%% file: main.tex
\documentclass[myepj-spec,pdftex]{mySvjour}
\usepackage{graphicx}
\usepackage[pdfpagemode=UseNone]{hyperref}
\usepackage{color}
\usepackage{xspace}
\usepackage[square,sort&compress,numbers]{natbib}   
\usepackage{verbatim}
\usepackage{amsmath,amssymb,amsfonts} 
\usepackage{tabularx}
\usepackage{multicol}
\usepackage{footnote}
\usepackage{listings}
\usepackage[gen]{eurosym}
\usepackage[switch, modulo]{lineno}

\usepackage{lmodern,bm}                
\usepackage[T1]{sansmath} 
\SetMathAlphabet{\mathsfbf}{sans}{\sansmathencoding}{\sfdefault}{bx}{sl}
\usepackage{etoolbox}
\AtBeginEnvironment{sansmath}{}{}{}

\usepackage{lipsum}

\definecolor{darkblue1}{rgb}{0,0,.2}
\definecolor{darkblue}{rgb}{0,0,.2}
\definecolor{darkred}{rgb}{0.5,0,0}
\pagecolor{white} 
\color{black}     
\hypersetup{breaklinks=true, 
	colorlinks=true, 
	linkcolor=darkblue1, 
	menucolor=darkblue1, 
	urlcolor=darkblue1,
	citecolor=darkblue1,
	pdftitle={},
	pdfauthor={},
	pdfsubject={},
	pdfkeywords={},
	pdfproducer={}
}
\usepackage{siunitx}

\bibstyle{plain}
\input{./Tools/command_list}

\begin{document}
	
	\twocolumn[{%
		\begin{@twocolumnfalse}
			
			\begin{flushright}
				\normalsize
			\end{flushright}
			
			\vspace{-2cm}
			
			\title{\Large\boldmath NaNu: Proposal for a Neutrino Experiment at the SPS Collider located at the North Area of CERN}
			%
			\input{author_list}
			
			\abstract{%
Several experiments have been proposed in the recent years to study the nature of tau neutrinos, in particular aiming for a first observation of tau anti-neutrinos, more stringent upper limit on its anomalous magnetic moment as well as new constrains on the strange-quark content of the nucleon. We propose here a new low-cost neutrino experiment at the CERN North area, named NaNu (North Area NeUtrino), compatible with the realization of the future SHADOWS and HIKE experiments at the same experimental area.}	
	\maketitle
	\end{@twocolumnfalse}
}]

\tableofcontents

\section{Introduction}	

\begin{figure*}[htb]
\begin{minipage}[t]{10.5cm}
	\centering
	\vbox{\includegraphics[width=10.5cm]{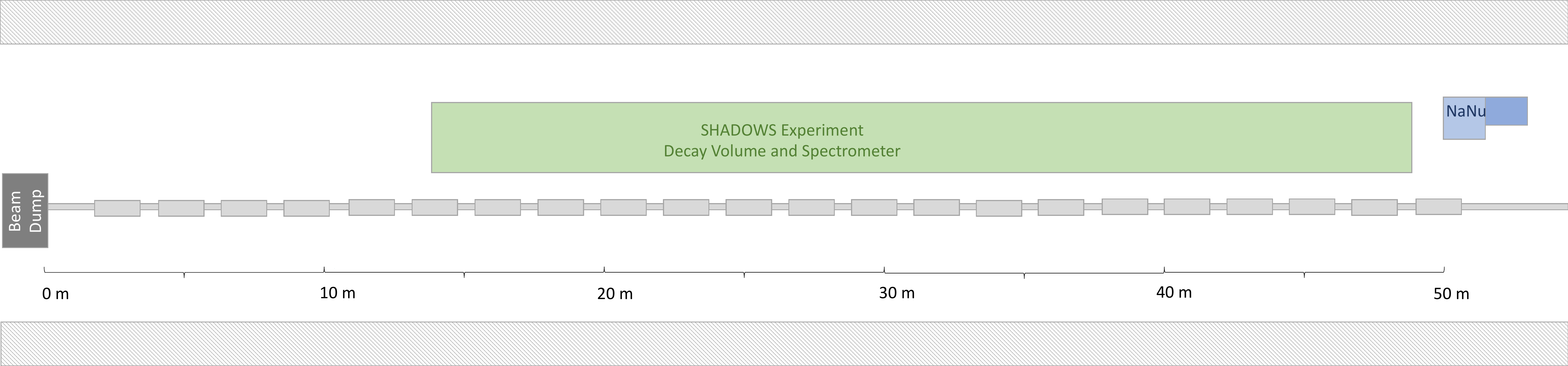}\\[0.3cm]}
	\caption{Schematic drawing of the top-view on the location of the NaNu Experiment in the CERN North Area together with the future SHADOWS Experiment and the beam-line.}
	\label{Fig:Location}
\end{minipage}
\hfill
\begin{minipage}[t]{6.0cm}
	\centering
	\includegraphics[width=6cm]{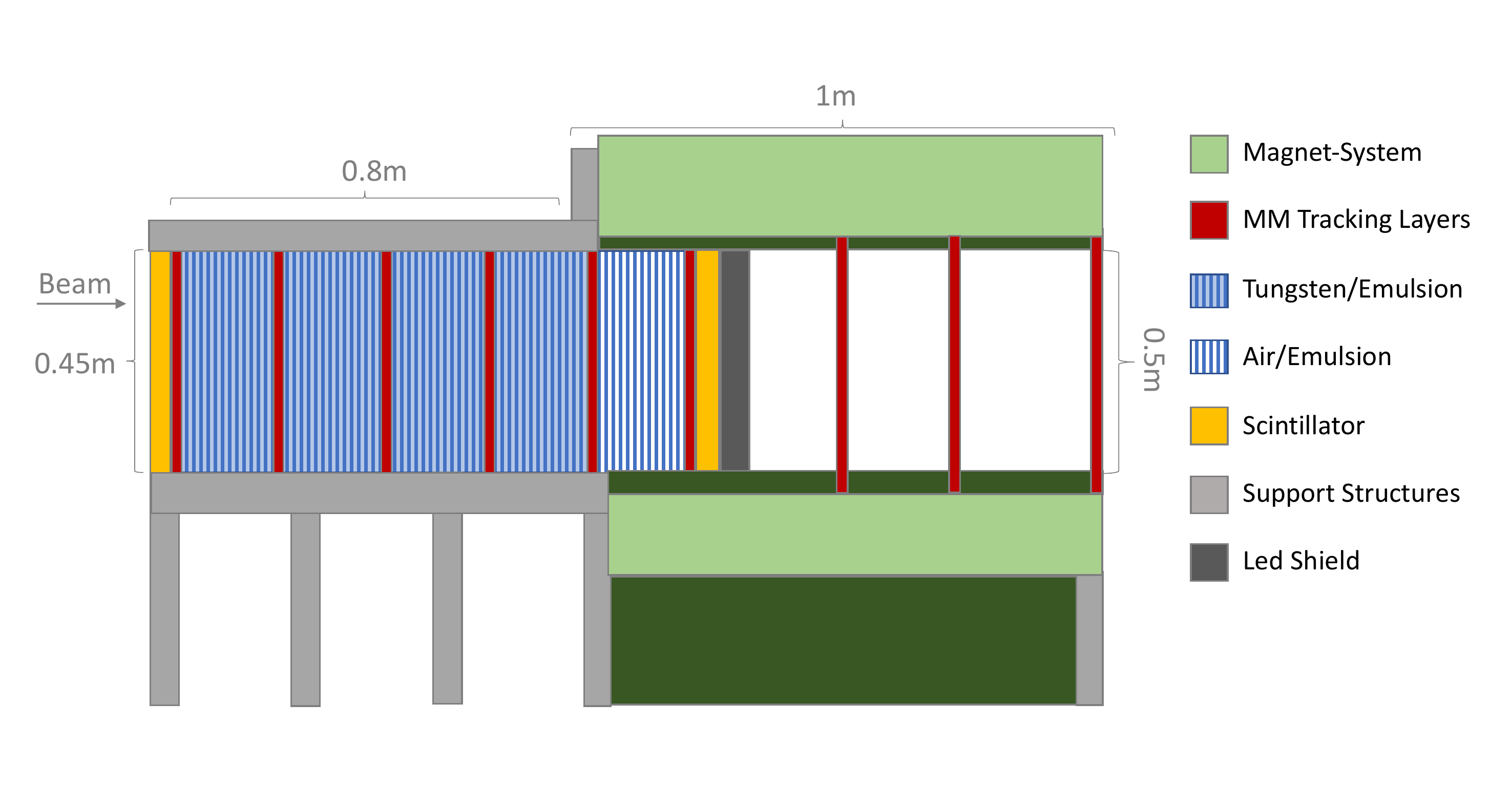}
	\caption{Side-view of the NaNu experiment with its major components: emulsion detector, magnet and tracking stations. The active detector is behind the emulsion detector and not shown.}
	\label{Fig:DetectorSchematic}
\end{minipage}
\end{figure*}

Within the SM, the neutrino sector is still the least understood and key questions, e.g. on the origin of the neutrino masses, are still not answered. Several new neutrino experiments are currently in preparation or just have started to take data. A particular interesting development are new experiments at colliders, e.g. FASER~\cite{FASER:2018eoc} and SND@LHC~\cite{Ahdida:2750060}, which aim for neutrino cross-section measurements in a new energy regime.  Of particular interest are tau neutrinos, since the study of their properties are limited to nine $\nu _\tau$ events observed at DONUT~\cite{DONUT:2000fbd} and ten $\nu _\tau$ candidate events at Opera~\cite{OPERA:2010pne}. The existence of anti-tau neutrinos has so far never been experimentally confirmed, making this the last missing particle within the SM. In fact, a huge number of tau- and anti-tau neutrinos could be produced in beam-dump experiments, where energies are high enough to produce $D_s^\pm$ mesons, which subsequently decay via $D_s \to \tau \nu_\tau$ with a branching fraction of about 5\%~\cite{Workman:2022ynf}. Originally, the SHiP collaboration suggested a dedicated neutrino detector to study such tau-neutrino events, with a convincing physics case~\cite{SHiP:2015vad, Ahdida:2654870, Bai:2018xum}. In this work, we propose a cost-efficient alternative neutrino detector, NaNu (North Area NeUtrino detector), which can be realized in-between the future SHADOWS and the HIKE experiments, as schematically shown in \autoref{Fig:Location}. The planned SHADOWS experiment~\cite{Baldini:2799412} at CERN is one possible realization of a beam dump experiment, located at the CERN North Area next to the SPS collider aiming for the search of dark matter and hidden particles. The SHADOWS detector is about 35~m long and 2.5~m wide, placed about 1~m off-axis and 14~m after the beam dump itself, where a 400~GeV proton beam provides $5\times10^{19}$ protons on target during a 4-year data-taking period between 2028 and 2032. The concept of SHADOWS foresees also the realization of the HIKE Experiment~\cite{HIKELoI}, which will be located about 50~m downstream of SHADOWS and will study extremely rare kaon decays. A first setup towards the full NaNu detector could be already installed in 2024 and operated together with the NA62 Experiment. 

In this work, we first discuss a preliminary NaNu detector concept, followed by an estimate on the neutrino fluxes, the identification of neutrino signatures as well as the physics reach and a cost estimate.

\section{Detector Concept}

The NaNu detector concept aims for a cost-efficient design using existing components and well established technologies. Depending on the available funding and the effectiveness of the background shielding, it can be realized in three different phases. We start with baseline concept in the following. 

A schematic drawing of the side-view of the standard NaNu detector is shown in \autoref{Fig:DetectorSchematic} with its four major components, namely the magnetic system, the emulsion target, the active trigger system as well as the muon spectrometer. For the baseline NaNu detector concept, we foresee to have two same-sized main detector components, labelled as \textit{active-} and \textit{emulsion-}detector in the following, both with dimensions of $45 \times 45\times 100$~cm$^3$. Both detectors are placed partly with a length of 20cm inside an existing dipole magnet at CERN with gap dimensions of $50 \times 100 \times 100$~cm$^3$ and a magnetic field strength of 1.4~T generated by a current of 2500~A. The transverse plane of the NaNu experiment, facing the interaction point has therefore a total size of $50 \times 100$~cm$^2$. 

The emulsion detector concept follows largely the current design of the FASER$\nu$ Experiment~\cite{FASER:2019dxq}. It consists of silver bromide crystals with diameters of 200~nm dispersed in gelatin media interleaved with a repeated structure of absorber plates in which the actual charged current neutrino interactions occur. Emulsion detectors have a spatial resolution between 50 to 100~nm and can be interpreted as detectors with a huge density of active channels, i.e. $10^{14}$ per cm$^3$. They are therefore perfectly suited for the study of short-lived particles with unique decay vertex structures. We propose to adopt the emulsion detector design of the successfully running FASER$\nu$ experiment, i.e. use emulsion films composed of two layers with a 70~$\mu$m thickness which are separated by a 200~$\mu$m thick plastic base. The emulsion films are interleaved by 1 mm tungsten plates due to their short radiation length. In total 560 tungsten plates with a total weight of the $\approx 2.2$~t are foreseen. Given the significant multiple scattering effects within the tungsten plates, we foresee to place 40 emulsion tracking layers in the remaining 20~cm within the magnetic field, allowing for momentum measurements with higher precision. Those layers will be stabilized by thin aluminium plates, that are separated by 4 mm air gaps. Since emulsion detectors cannot record timing information, all charged particles leave tracks and lead to significant pile-up. The emulsion detector is therefore complemented by six micromegas based tracking detectors with two-dimensional readout as well as a two-gap design, originally proposed in~\cite{Brickwedde:2016lhu}. The thickness of these active detectors is about 15~mm and yields a spatial resolution of $\approx 150$~$\mu$m in two spatial dimensions with 1800 readout channels for each detector. The two-gap design allows in addition the reconstruction of complete tracklets with angular resolution of 0.03~rad. The emulsion detector is designed to identify electron-, muon- and tau-neutrino interactions. 

The active detector has the same dimensions as the emulsion detector. It consists of 2.6cm thick tungsten plates, interleaved with 0.9mm thick plastic-scintillators with a SiPM readout system with ten channels on each layer. Similar to the emulsion detector, one Micromegas tracking layer is placed every 15cm. The total weight of the tungsten is $\approx 2.5$~t. While the Micromegas tracking layers are foreseen to measure the angle of transversing muons, the plastic-scintillators are used to measure hadronic shower energies. The active detector is therefore a combination of a tracking detector and a hadronic sandwich calorimeter. While it cannot identify electron- and tau-neutrino interactions, it is perfectly suited to measure interactions of muon neutrinos, in particular the angle of the muon as well as the energy of the hadronic recoil system. 

Before and after both detector systems as well as on the side facing the beam-line, a highly efficient muon-veto system based on plastic-scintillators is foreseen, to reduce the background of muons for the active components of NaNu.  
 
The emulsion detector and active detector are followed by a muon spectrometer consisting of four layers of the same micromegas-based technology as previously described but with larger dimensions of $100\times 50$~cm$^2$ with 3600 readout channels per detector layer. The four layers are separated by 20~cm each, while the first layer is shielded in addition with a 20~cm iron layer to suppress hadronic particles. The active trigger system of NaNu makes use of plastic scintillators, which are located in the front as well as in the back of the emulsion layers, where they can be also used to veto muon signatures. In addition, the self-triggering capabilities of micromegas detectors can be used to record events which only leave signatures within the emulsion target. 

The location of the NaNu Experiment is foreseen 50~m after the beam dump, i.e. behind the SHADOWS experiment. The distance to the beam axis is chosen to minimize the expected muon background. Muons for the SHADOWS experiment are shielded by dedicated magnetized iron blocks. The expected muon background for $4\times10^{19}$ proton on target in the transverse plane to the beam line is shown in \autoref{Fig:Background}. The neutrino flux and the neutrino energies increase when moving towards the beam-line, however, also the muon-flux increases significantly. It is typically assumed that the reconstruction algorithms of emulsion detectors can handle $\approx 10^6$ tracks per cm$^2$. The position of NaNu emulsion detector in the transverse plane was therefore optimized to be closest as possible to the beam-line, but the number of $10^6$ muons per cm$^2$ per six months operation is not exceeded. This implies a potential location of the NaNu magnet gap in the transverse plane with the coordinates $x_1$ = 0.55m to $x_2$ = 1.55m and $y_1$ = -0.2m and $y_2$ = 0.2m, where the $y=0m$ corresponds to the center of the beam dump. The active detector component is foreseen between $x_1$ = 0.55m to $x_2$ = 1.0m, while the emulsion detector would be placed between $x_1$ = 1.0m to $x_2$ = 1.45m in the baseline NaNu experiment.

\begin{figure*}[thb]
\centering
    \includegraphics[width=0.32\textwidth]{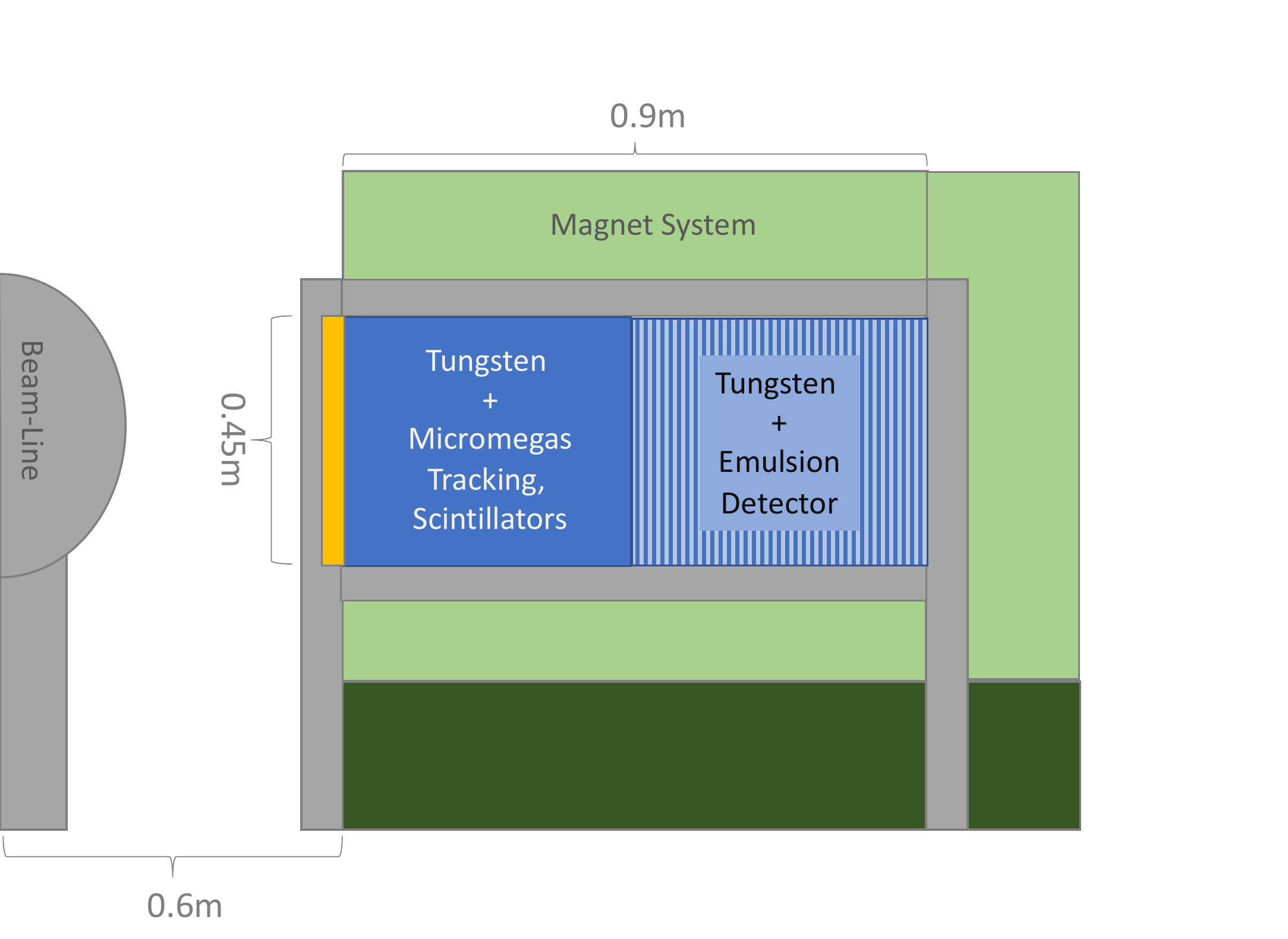}%
    \includegraphics[width=0.32\textwidth]{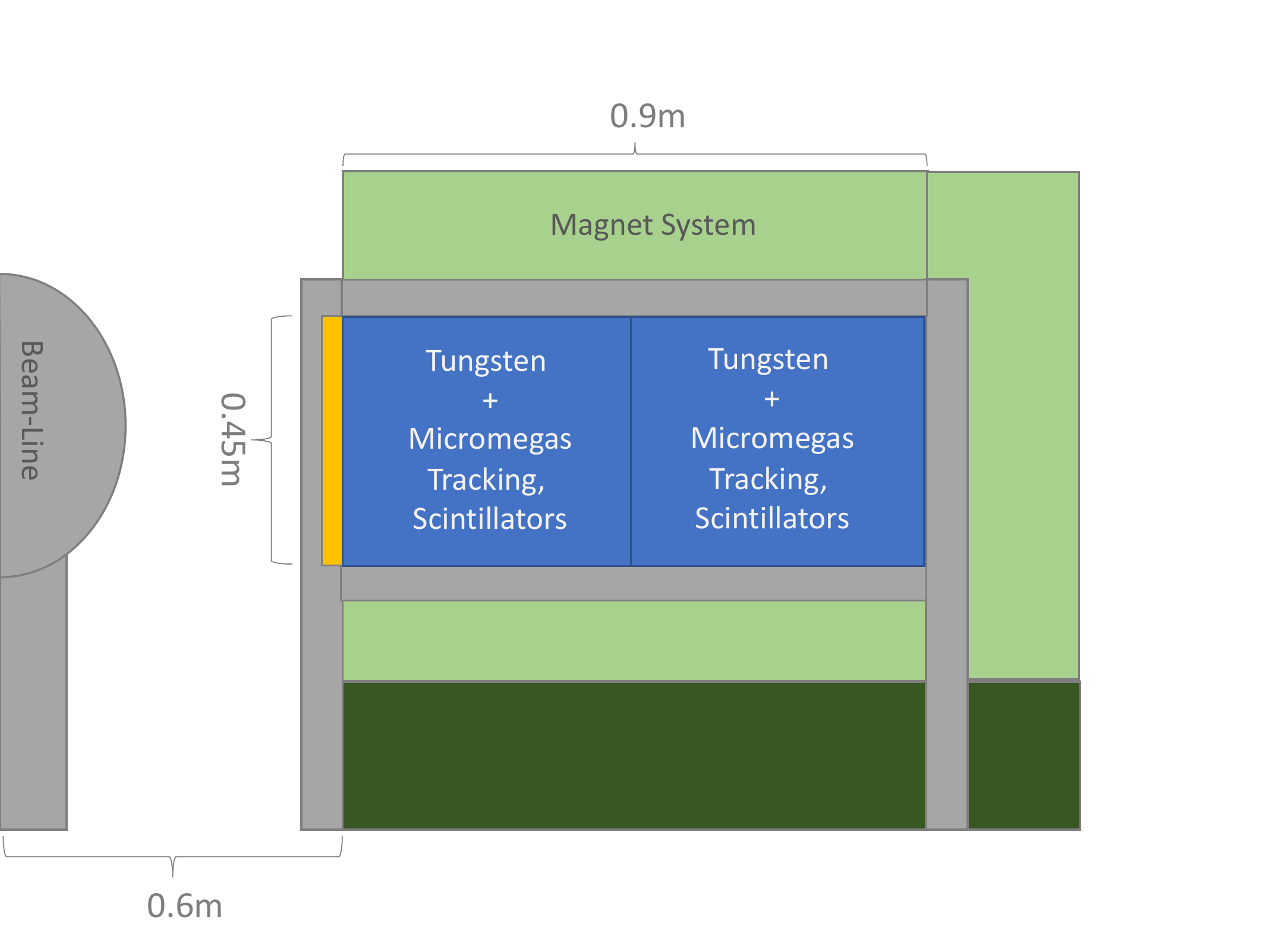}%
    \includegraphics[width=0.32\textwidth]{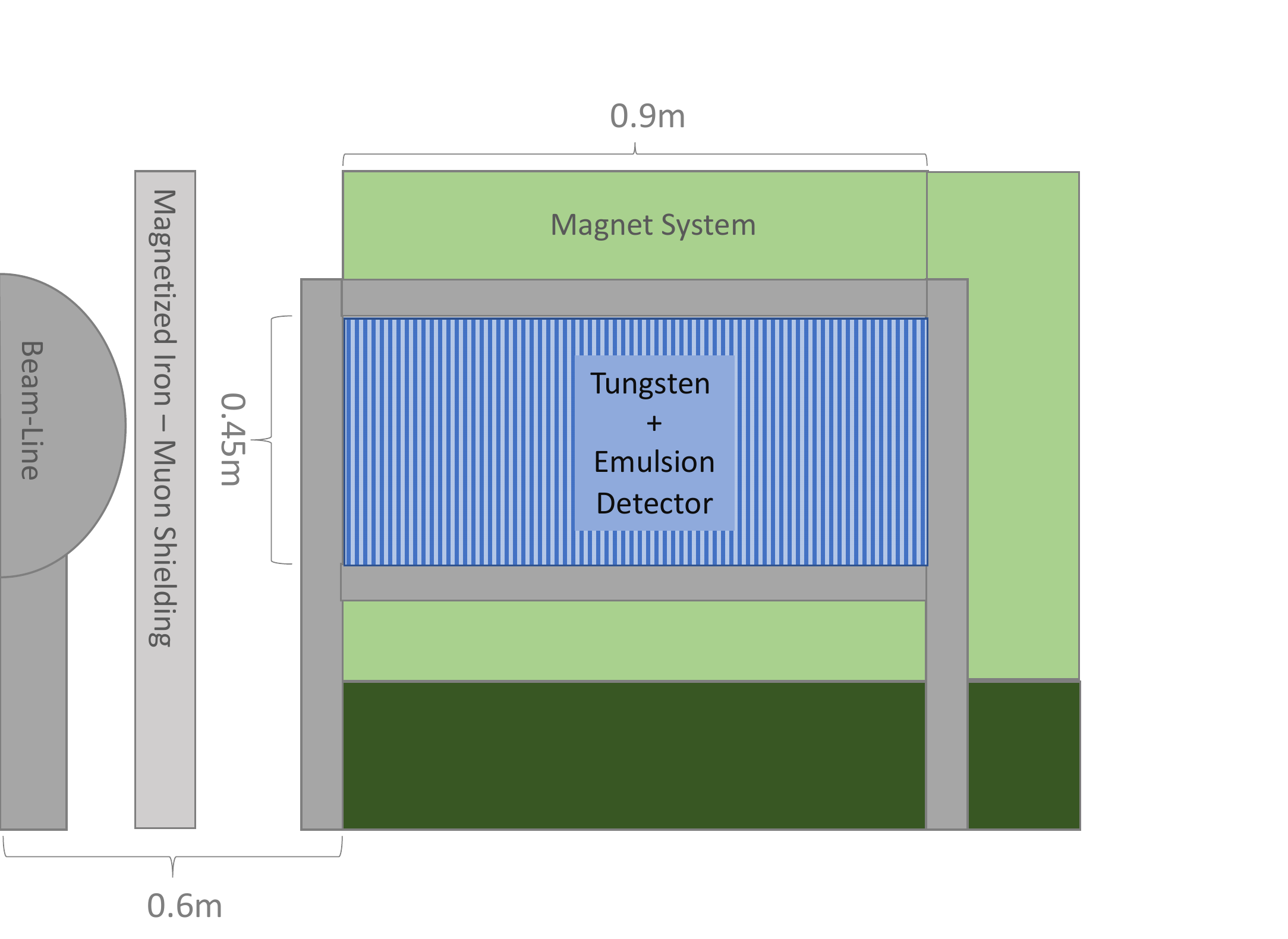}%
\caption{Front-view of the three different stages of the NaNu experiment: The baseline NaNu detector (left) has one emulsion-based detector as well as one active detector based in micromegas and scintilator technologies; the Mu-NaNu detector (middle) omits the emulsion detector and uses only active detector components; the Super-NaNu detector consists of a full emulsion based-detector concept (right).}
\label{fig:FrontView}
\end{figure*}

The NA62 experiment, currently taking data at the future location of SHADOWS and HIKE, runs also partly in beam-dump mode. A first version of the NaNu experiment, named Mu-NaNu, could be already installed in the coming years, placed at a similar location as the future baseline NaNu experiment. The detector concept of Mu-NaNu would be largely equivalent to baseline NaNu detector concept, however, the emulsion detector would be replaced by a second active detector block, as illustrated in front-views of the NaNu experiment in Figure \ref{fig:FrontView}. The Mu-NaNu detector is therefore perfectly suited to reconstruct muon neutrino interactions and study the corresponding deep inelastic scattering processes. Even more importantly, it would serve as an ideal platform to test and study technologies for the later experimental stages.
 
The active detector component can be replaced by a second emulsion detector, depending on the final background rate of muons, the effectiveness of a dedicated muon shielding as well as the availability of sufficient funding. This setup, Super-NaNu, would drastically increase the rate of recorded electron and tau neutrinos, as it increases the material-budget of emulsion detector by a factor of two and - even more importantly - moves the detector closer to the beamline, where higher neutrino rates and energies are expected. We will discuss the potential physics reach in the following.

\section{Neutrino Fluxes}	

Tau-neutrinos passing through the NaNu detector stem primarily from mesonic decays of $D_s\to \tau \nu_\tau$,  together with a high rate of $\nu_\mu$ and $\nu_e$ neutrinos, which appear as decay products of charmed hadrons and soft pions and kaons. Our estimate of neutrino fluxes is based in studies of \textsc{Pythia8} ~\cite{Bierlich:2022pfr}  and the \textsc{Genie} program~\cite{Andreopoulos:2009rq} and then cross-checked with the published studies of the Scattering and Neutrino Detector (SND) of the SHIP experiment~\cite{Alekhin:2015byh,SHiP:2015vad,Ahdida:2654870}. The SND detector is assumed to be placed about 35~m after the interaction point, directly following the beamline. Its neutrino interaction target has a transverse size of $0.8 \times 0.8$~m$^2$, a depth of $\approx 1$~m and consists of passive absorber material interleaved with 19 emulsion and 19 tracking planes. 

Since the production processes of neutrinos is the same for SND as for NaNu, the kinematic distributions of the produced neutrinos is equivalent for both experimental setups. However, the differences of the expected number of neutrino interactions are due to the different location, the different size as well as the different assumed number of protons on target.

\begin{figure*}[htb]
\begin{minipage}[t]{\columnwidth}
	\centering
	\includegraphics[width=\textwidth]{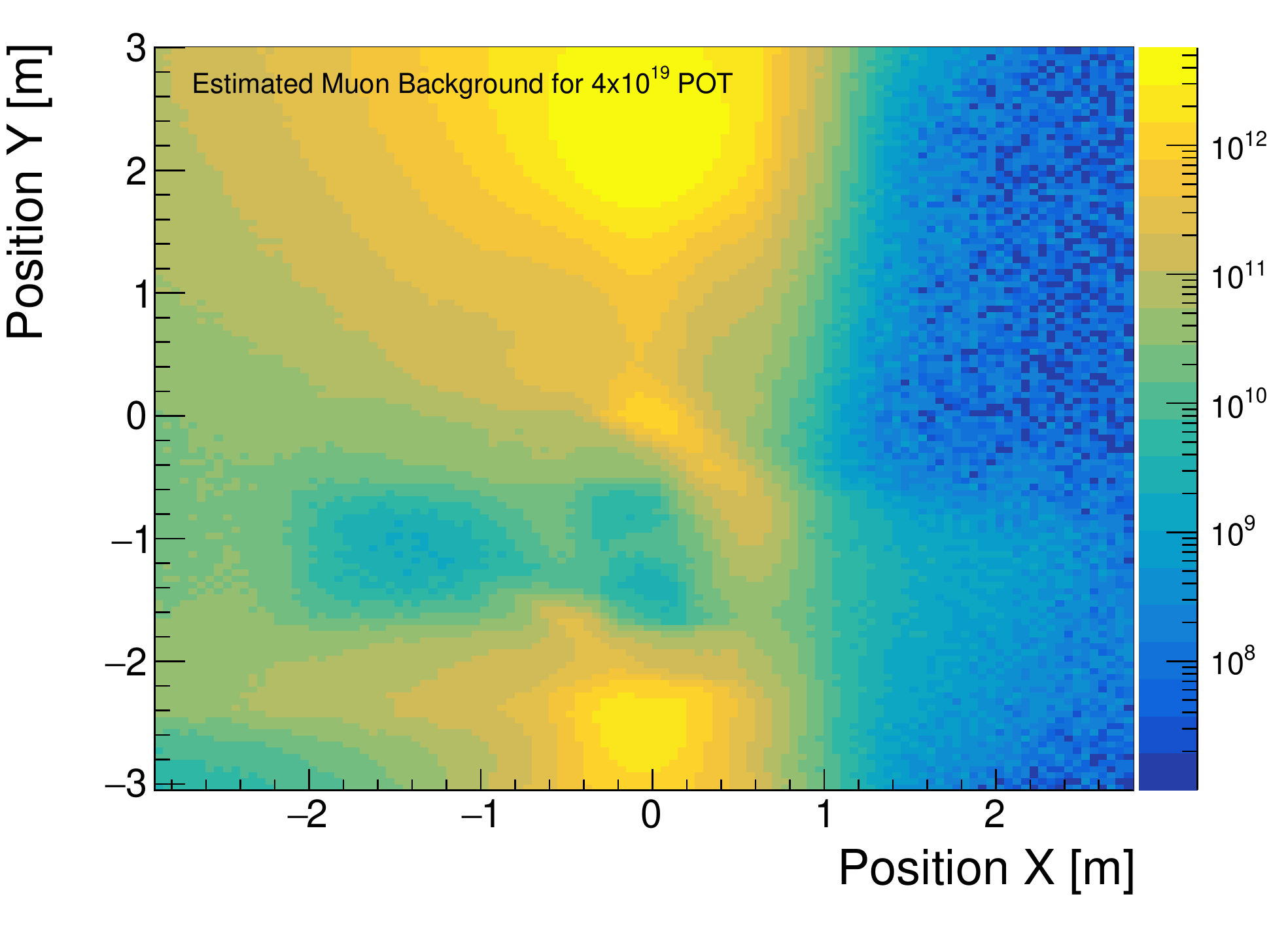}
	\caption{Estimated muon background in the transverse plane to the beam axis at a distance of 50m after the beam dump for the expected $4 \times 10^{19}$ protons on target.}
	\label{Fig:Background}
\end{minipage}
\hfill
\begin{minipage}[t]{\columnwidth}
	\centering
	\includegraphics[width=\textwidth]{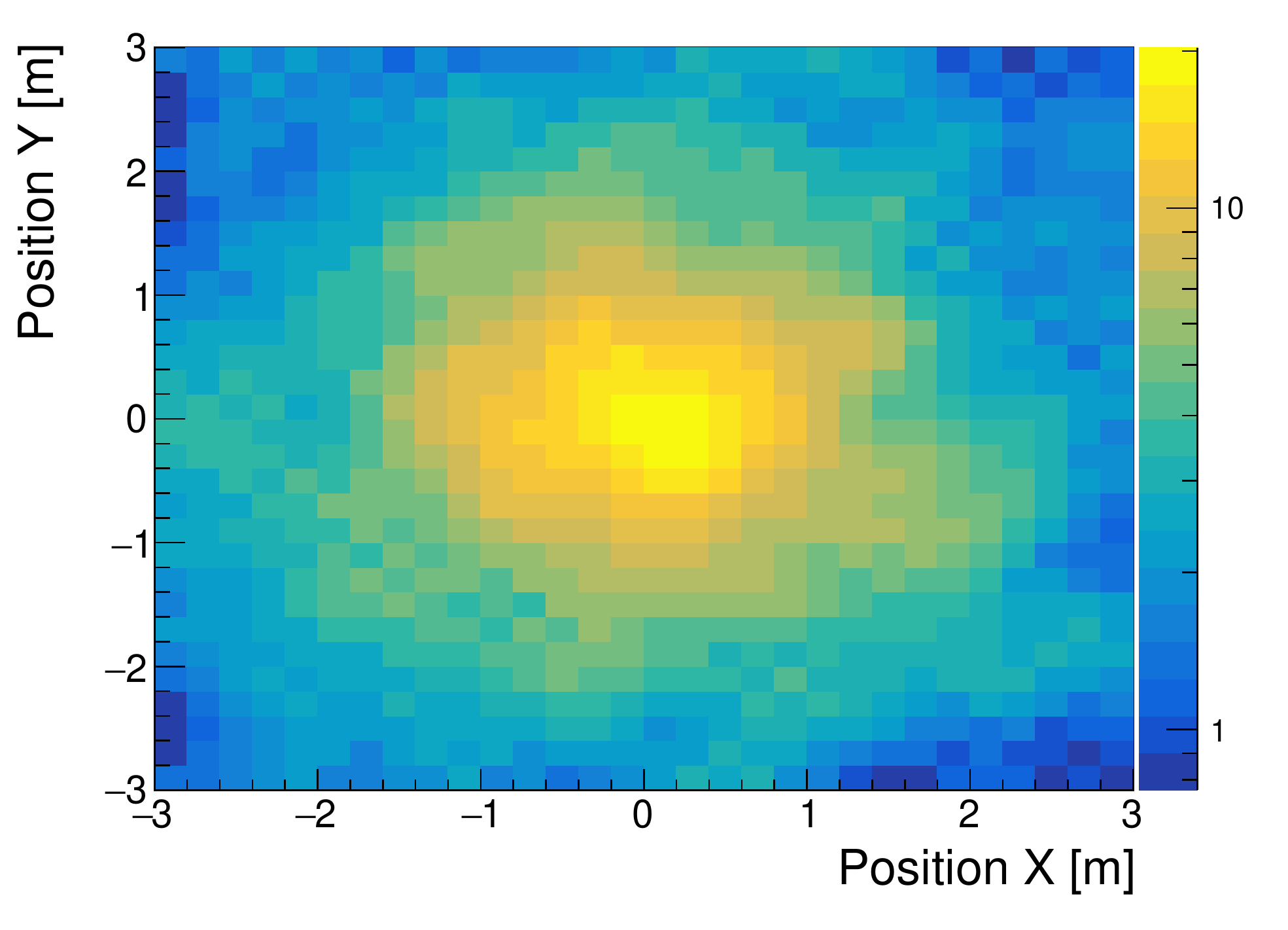}
	\caption{Distributions of $\tau$ neutrinos produced in 400~GeV proton beam collisions on a fixed proton target using \textsc{Pythia8} 50m after the beam dump.}
	\label{Fig:NeutrinoFlux}
\end{minipage}
\end{figure*}

In four years of operation, we assume to have collected data corresponding to $5\times10^{19}$ POT, i.e. a factor of 4 less compared to SND assumptions. The amount of passive absorber material 
also is significantly smaller for the baseline NaNu experiment with a volume of 0.11~m$^3$ and a weight of 2.2~t, yielding a further reduction by a factor of 3.3. Most important, however, is the difference in the actual location.
\textsc{Pythia8}~\cite{Bierlich:2022pfr} was used to estimate the kinematic distributions of all neutrino flavours after the initial proton-proton interactions. The expected spatial distributions for tau neutrinos 50m after the beam dump is shown in \autoref{Fig:NeutrinoFlux}. In addition, the neutrino energy dependence on the neutrino interaction cross-sections have to be considered. In fact, the average neutrino energy reduces by up to 30 GeV, depending on the neutrino flavour, when placing the emulsion target of the NaNu Detector with a distance of 1.0 to 1.5~m to the beam axis, implying significant lower cross-section in comparison to more high energetic neutrinos directly in the beamline. The expected neutrino interactions in the NaNu detector due to its position is expected to be therefore reduced by a factor of 20. 

While the Super-NaNu setup has also a similar target mass compared to the baseline NaNu detector, it would allow for a study of electron- and tau-neutrino interactions closer to the beam-line with a  emulsion detector that covers twice the target mass. Hence the expected reduction factor for the mass is only 0.6 with respect to SND. Moving closer to the beam-pipe, increases the neutrino rate but also increases their energies. Given the large cross-section dependence on the neutrino energies, an increase in the number of neutrino interactions within NaNu by a factor of two to three is expected (Table \ref{tab:neutrinoflux}). Moreover, we expect a slight difference in the overall increase of flux between neutrinos and anti-neutrinos due to their different production and interaction cross sections. In should be noted that no increase in the muon neutrino flux is expected, since we assume that muon neutrino interactions can be reconstructed in the active detector component as well as the emulsion detector and no change in the overall target mass between the Super-NaNu and the baseline setup is foreseen.

The installation of the Mu-NaNu detector possibly in the same position as the baseline NaNu detector, with data-taking from 2024 to 2026 could result in $0.5-1.0\times 10^{18}$ POTs, with a similar target mass as the baseline NaNu detector, yielding 700 muon neutrino and 150 anti-muon neutrino interactions. Given that Mu-NaNu can only reconstruct muon neutrino interactions, no study of electron and tau-neutrino events is foreseen. 

The final estimates of expected neutrino interactions per flavor are summarized in \autoref{tab:neutrinoflux} for the three setups of NaNu. 

\begin{table}[ht]
\begin{center}
\begin{tabular}{l | c | c | c}
\hline
Experimental 		& Mu-NaNu		& NaNu				& Super-NaNu \\
Setup			& 				& 					& 			 \\
\hline
$\nu_e$		    	& -				& $4.1 \times 10^3$		& $20 \times 10^3$ \\
$\bar \nu_e$		& -		 		& $1.0 \times 10^3$		& $4.5 \times 10^3$ \\
\hline
$\nu_\mu$		& 700 			& $40 \times 10^3$		& $40 \times 10^3$\\
 $\bar \nu_\mu$		& 150 			& $9 \times 10^3$ 		& $9 \times 10^3$\\
\hline
$\nu_\tau$	    	& - 				& $0.12 \times 10^3$ 	& $0.72 \times 10^3$ \\
$\bar \nu_\tau$		& - 				& $0.07 \times 10^3$	& $0.41 \times 10^3$ \\
\hline
\end{tabular}
\end{center}
\caption{Expected number of detectable neutrino interactions within the NaNu detector for $4\times10^{19}$ POT for NaNu and Super-NaNu as well as for $1\times10^{18}$ POT for Mu-NaNu.}
\label{tab:neutrinoflux}
\end{table}

\section{Detector Simulation and Neutrino Identification}
The NaNu detector concept has been implemented in \textsc{Geant4}~\cite{Agostinelli:2002hh} while the primary neutrino interactions with the passive detector material have been simulated via the  \textsc{Genie} program. It should be noted that the theoretical uncertainties on the actual neutrino fluxes as well as the hadronization model are typically larger than the experimental uncertainties involved in the following studies. 
 
Neutrino interactions of all flavours require at least two tracks with an energy of 1~GeV associated to a common vertex with an impact parameter smaller than 10~$\mu$m, one stemming from the lepton, the other(s) from the recoil of the nucleus. Tracks with larger impact parameters are treated as seed candidates for a secondary vertex. Secondary vertices can be defined via kinks on their corresponding tracks w.r.t. to the primary tracks as well as a by displacement of more than 20~$\mu$m to the primary vertex. Primary and secondary decay vertices are searched for in all emulsion layers of NaNu except those close to the downstream tracking layers with a 1~cm margin. This reduces the acceptance of the NaNu detector by approximately 5\%.

\begin{figure*}[thb]
\centering
    \includegraphics[width=0.495\textwidth]{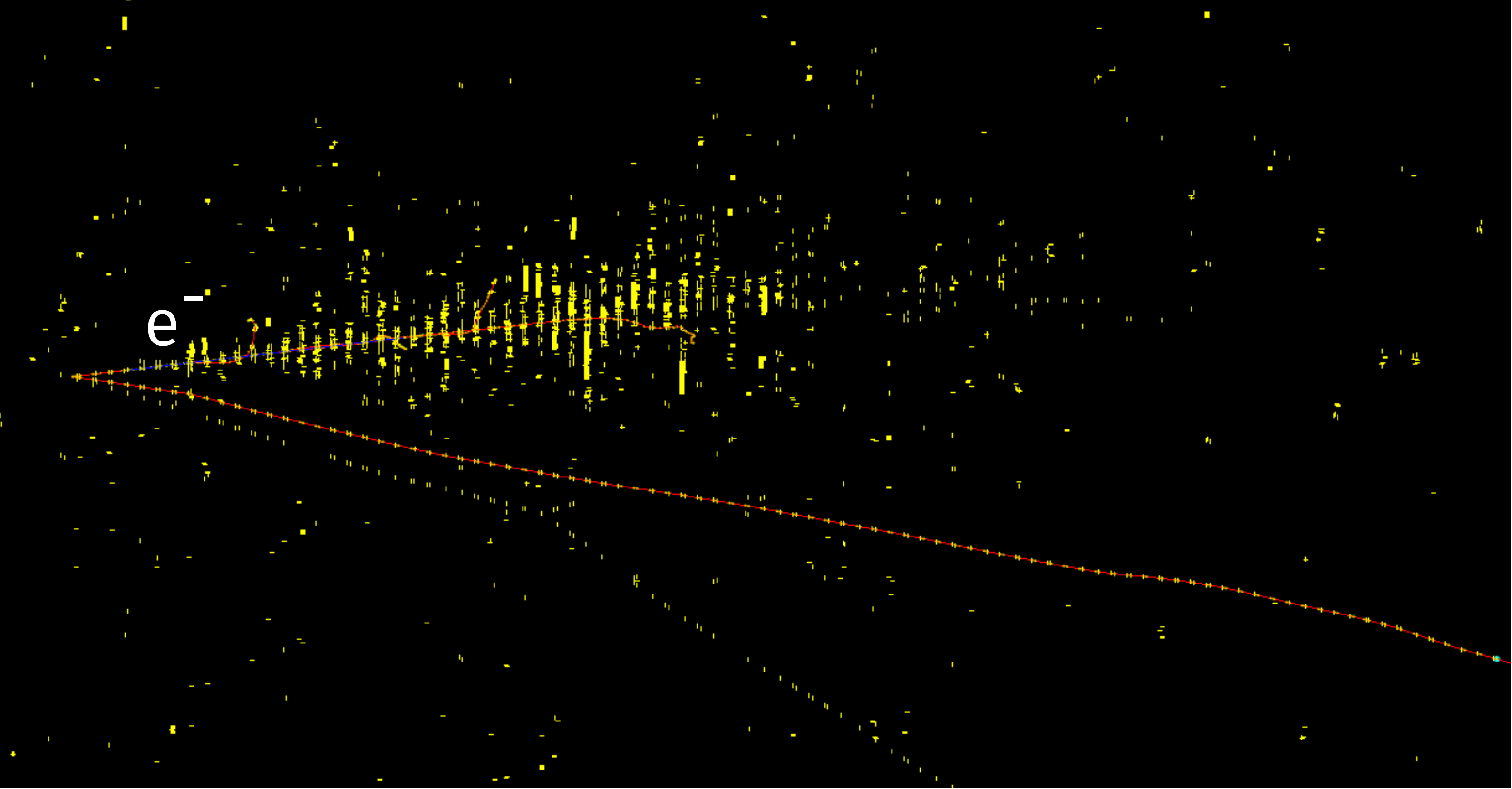}%
    \hfill\includegraphics[width=0.495\textwidth]{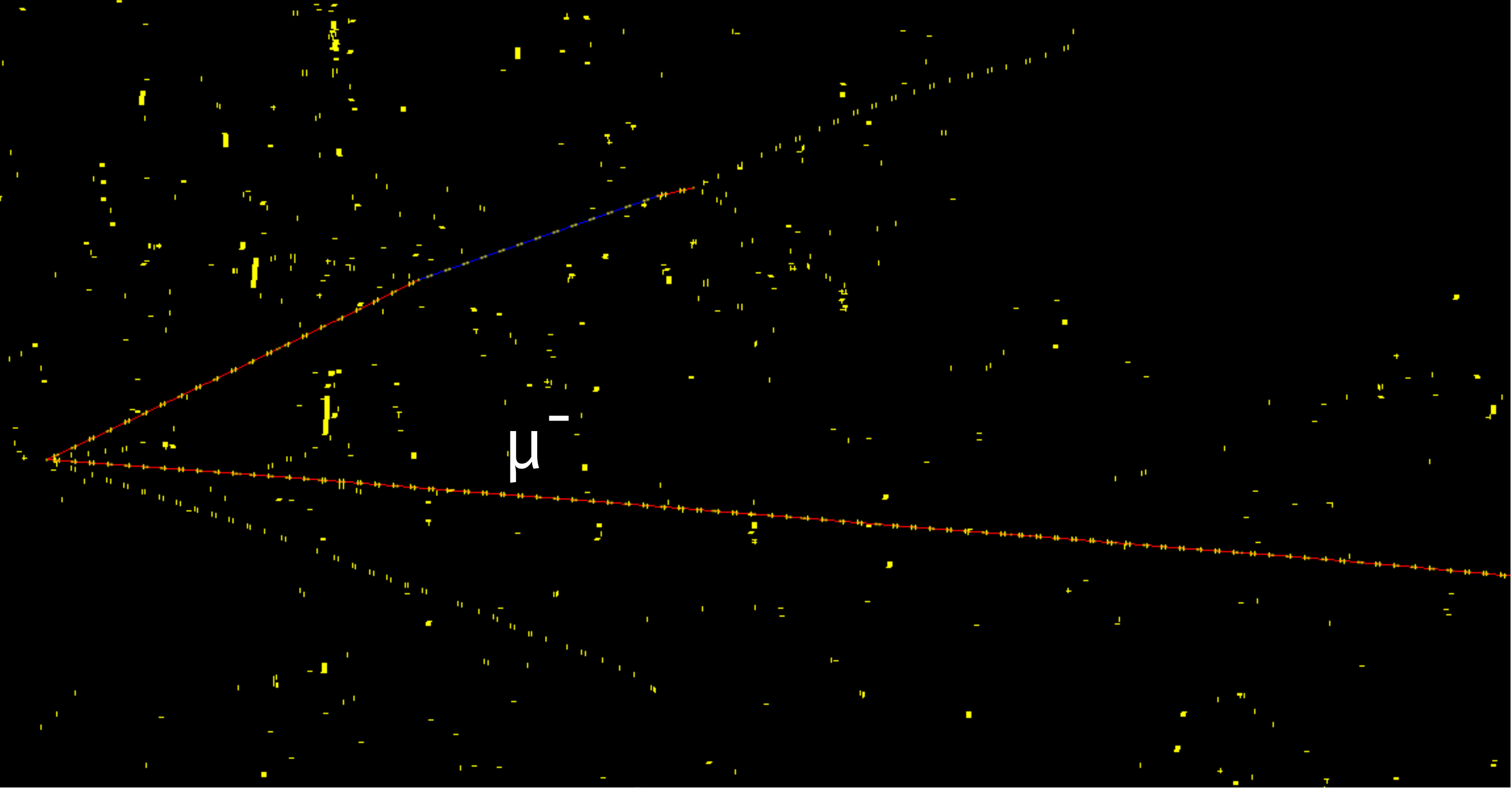}\\
    \includegraphics[width=0.495\textwidth]{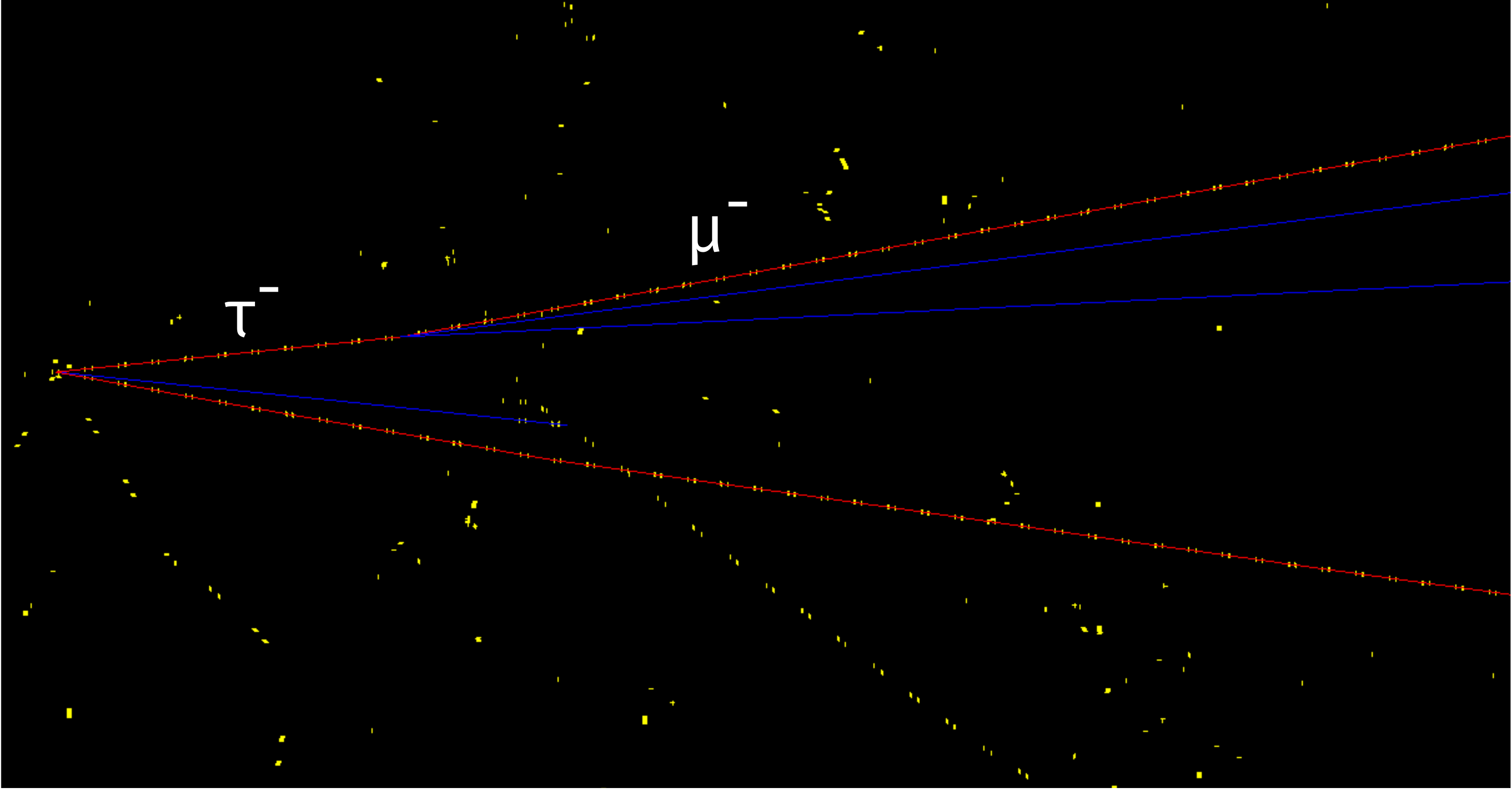}%
    \hfill\includegraphics[width=0.495\textwidth]{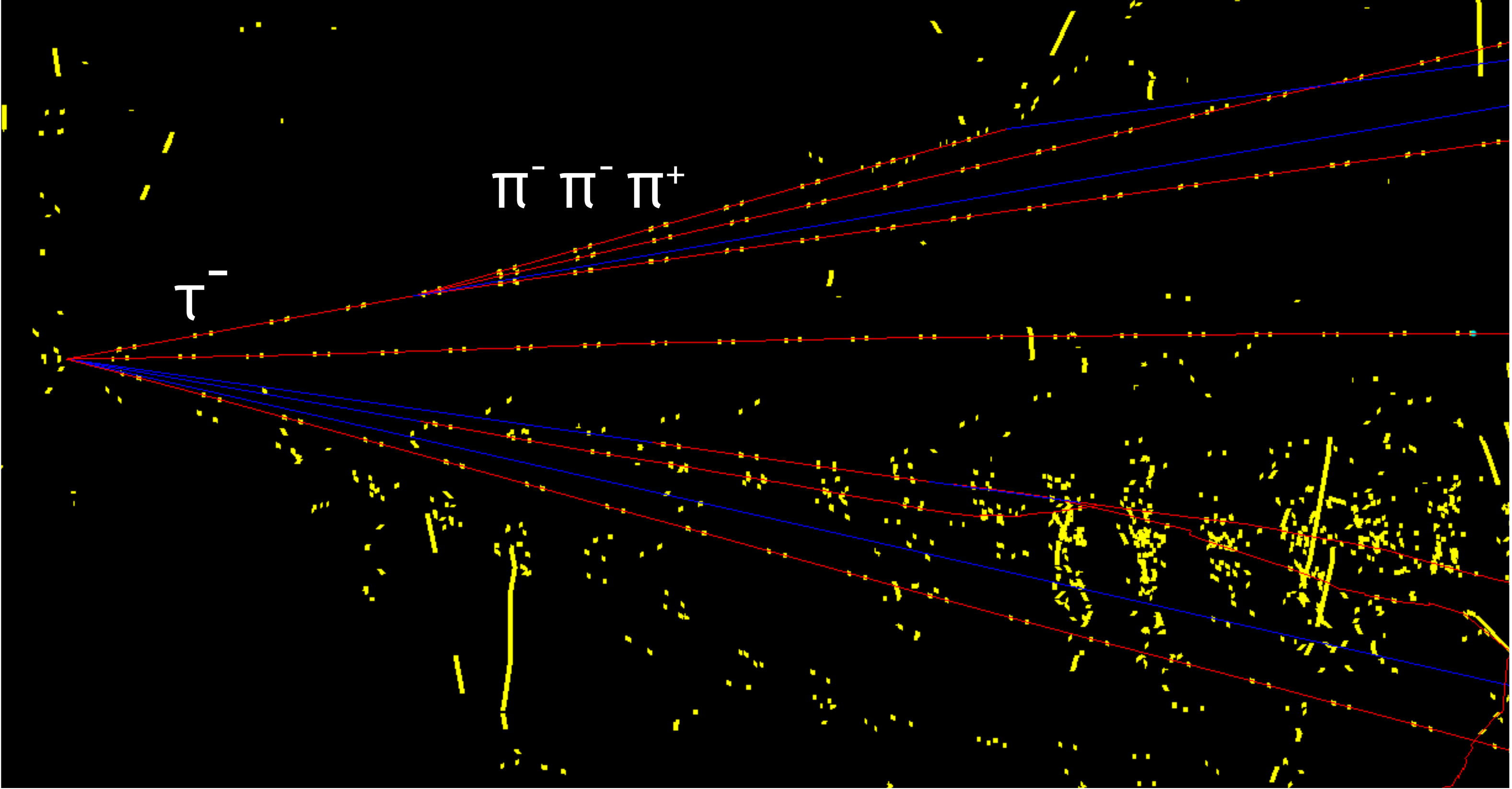}
\caption{Event Displays of the primary and secondary particle trajectories of electron neutrino (upper left) and muon neutrino (upper right) interactions with the NaNu Detector. Two example of tau neutrino interactions are shown in the lower row in the muonic decay channel (left) and the $\tau\rightarrow \pi^+\pi^-\pi^-$ decay channel (right). The charged (neutral) particles are indicated by red (blue) lines. Only particles with momenta larger than 1 GeV are shown.}
\label{fig:EventDisplays}
\end{figure*}

Electron neutrino interactions are identified by the characteristic electromagnetic showers of the electron as shown in \autoref{fig:EventDisplays}. A 90\% detection efficiency is assumed, in line with studies of similar neutrino experiments. Given the short electron tracks, no charge measurement is possible. In total, we expect about \num{4600} reconstructed electron neutrino events in the baseline NaNu detector, while \num{22000} in the Super-NaNu setup. 

Muon neutrino interactions (\autoref{fig:EventDisplays}) in the emulsion detector can be identified by a primary vertex in the emulsion layer with one long track leaving the vertex. However, the active detector component is specifically designed to study muon neutrino interactions. The angle of the muon-track is reconstructed with the Micromegas layers within the target and we expect a relative angle resolution between 0.1 and 0.8 degrees for neutrino energies between 10 and 60 GeV, limited by multiple scattering effects for low energetic muons. The expected muon background rate poses no problems for Micromegas detector technology, which has proven to work reliably in high-rate environments. 

\begin{figure*}[htb]
\begin{minipage}[t]{\columnwidth}
	\centering
	\includegraphics[width=\textwidth]{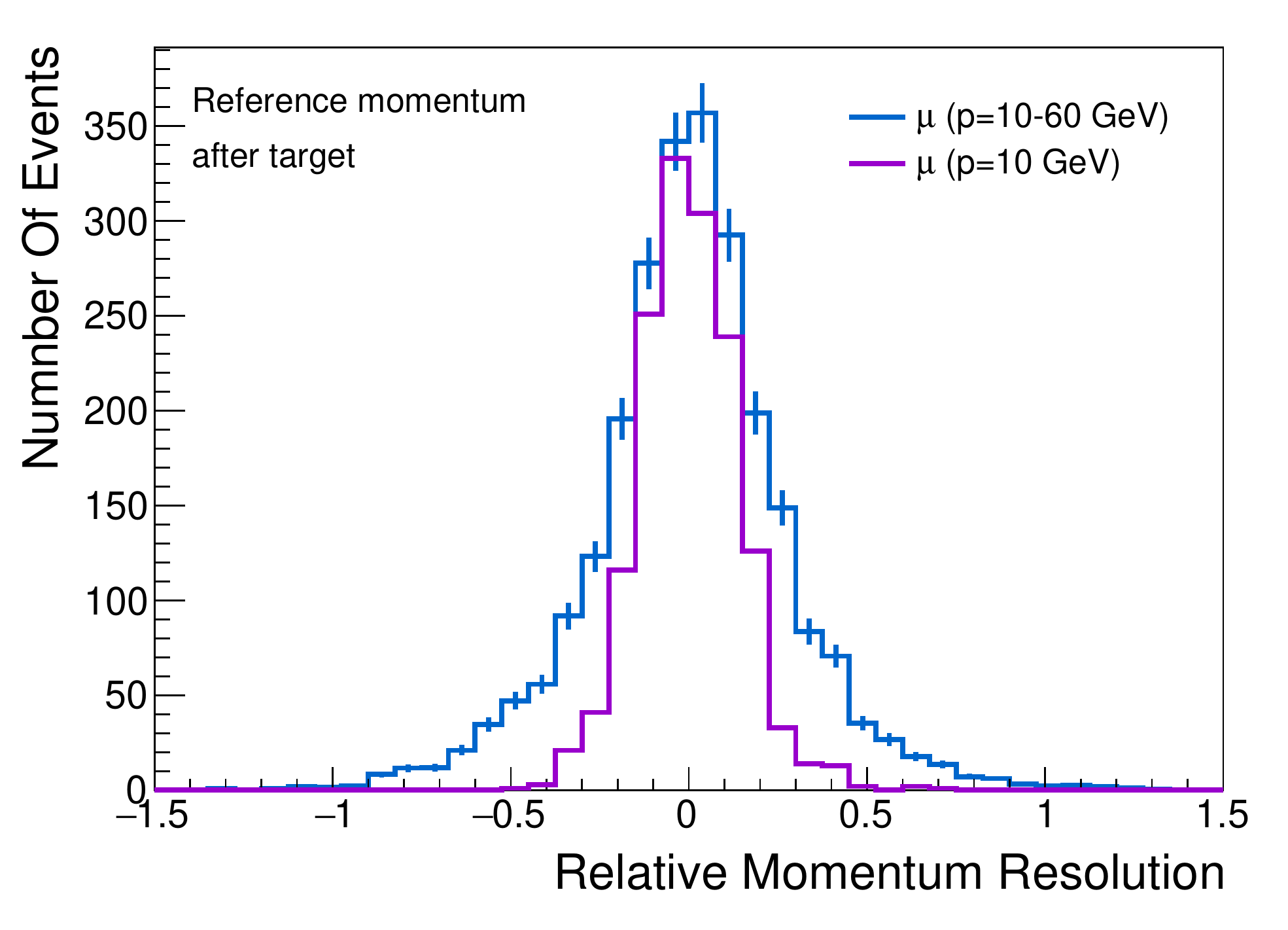}
    	\caption{Relative momentum resolution for muons with an energy of 5 and 10 GeV.}
   	 \label{fig:MomentumRes}
\end{minipage}
\hfill
\begin{minipage}[t]{\columnwidth}
	\centering
	\includegraphics[width=\textwidth]{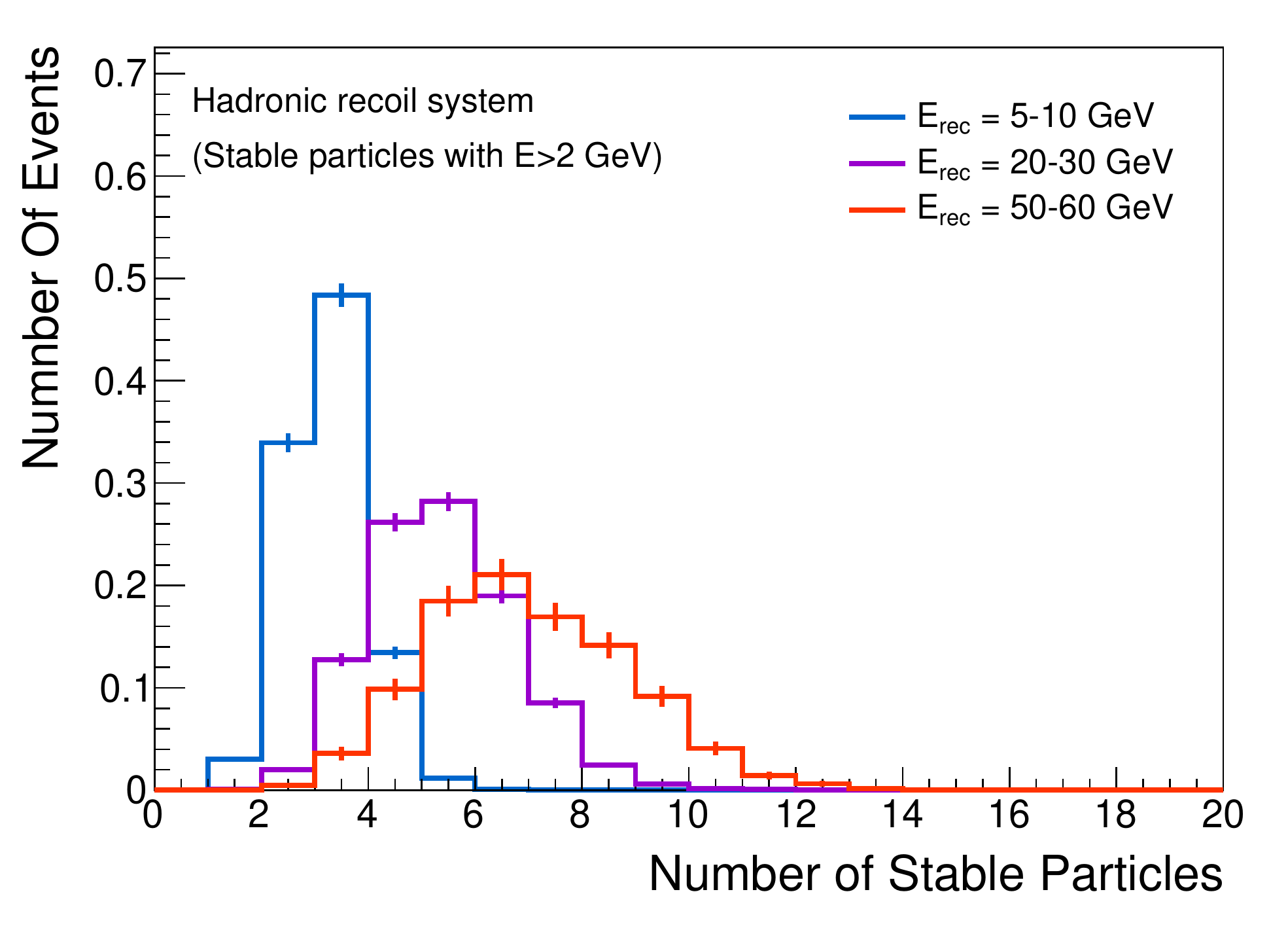}
	\caption{Hadronic.}
	\label{fig:Hadron}
\end{minipage}
\end{figure*}

The curvature of the track in the muon spectrometers allows for a momentum measurement. The expected momentum resolution for muon tracks in the NaNu detector has been estimated using a full \textsc{Geant4} simulation. Figure \ref{fig:MomentumRes} shows the expected momentum resolution for the inclusive muon spectra as well as for muons with an energy of 10 GeV, comparing to the muon momentum after the target. An energy loss in the order of 1-2 GeV of muons in the target needs to be corrected in the final momentum estimate. Given the similar size of the magnet-gap and the active detector system, acceptance losses of up to 30\% are expected for the muon identification in the muon spectrometer. In total, we expect \num{30000} and \num{6000} reconstructed muon neutrino and anti-muon neutrino events, respectively for the baseline and the Super-NaNu setups. The design of the active detector component as hadronic sandwich calorimeter allows not only for a vertexing of the final-state muon via the micromegas detectors, but also a measurement of the hadronic recoil energy using the scintillators. Given the significant dependence on the number of final state hadrons after a muon neutrino interaction (Figure \ref{fig:Hadron}), also the Micromegas layers can be used for the identification and the measurement of hadronic shower-shape variables. We expect an energy resolution on the hadronic recoil system of $200\%/\sqrt(E[GeV])$. Further optimization studies are currently ongoing.

The event topology of tau neutrino interactions depends on the subsequent decay of the tau lepton and, hence, the presence of a secondary vertex. The secondary vertex is expected to be reconstructed with an efficiency of $\approx 60$\%. Tau decays to electrons or muons yield a kink between the original tau track and the subsequent electron or muon tracks. The decay leptons can be identified as described above. The hadronic tau decays are identified by their short-lived decays and distinguished by the number of charged hadrons in the final state, i.e. into one or three charged pions. The three-pion decay has a unique topology, while the one-pion decay has a similar vertex structure as the muonic decay, however with a shorter track-length of 12.5~cm. This allows also for a momentum measurement using the tracking information within the magnetic field, where a significant better momentum resolution is achieved for tau-decays that happen close to end of the passive emulsion detector layers. 

The total number of reconstructed $\nu_\tau$ and $\bar \nu_\tau$ interactions at NaNu and Super-NaNu are about 100 and 570, respectively, as detailed in \autoref{tab:neutrinoreco} together with the efficiency losses at various stages of the identification. In the current setup, only the charge in the muonic tau decay channel can be measured. Neglecting a possible charge identification in other final states, we expect to identify the charge 10 $\nu_\tau$ and 7 $\bar \nu_\tau$ interactions at NaNu as well as 60 $\nu_\tau$ and 40 $\bar \nu_\tau$ interactions at Super-NaNu for $4\times10^{19}$ POT.

The main background in the tau lepton identification is due to the charm hadron production in $\nu _\mu$ interactions via charged current interactions in the passive detector material. The subsequent decay topology of those hadrons has a similar decay topology as the tau decays when the primary lepton is not correctly identified. Hence the performance of the muon identification system is crucial for the background rejection. Given a larger reduction of the $\nu_\mu$ interaction rate compared to SND@SHIP, we expect also a similar or even smaller background contribution in the $\nu_\tau$ identifications.

\begin{table*}[htb]
\centering
\begin{tabular}{l |cccc|cccc}
\hline
Decay-Channel		& $\tau\rightarrow e$	& $\tau\rightarrow \mu$	& $\tau\rightarrow h (\pi^\pm)$	& $\tau\rightarrow 3h (3\pi^\pm)$	&  $\bar \tau\rightarrow e$	& $\bar \tau\rightarrow \mu$	& $\bar \tau\rightarrow h (\pi^\pm)$	& $\bar \tau\rightarrow 3h (3\pi^\pm)$	\\
\hline
BR				& 0.17	& 0.18	& 0.46	& 0.12	& 0.17	& 0.18	& 0.46	& 0.12	\\
Geometrical		& 0.9		& 0.9		& 0.9		& 0.9		& 0.9		& 0.9		& 0.9		& 0.9		\\
Decay search 		& 0.6		& 0.6		& 0.6		& 0.6		& 0.6		& 0.6		& 0.6		& 0.6		\\		
PID 				& 1.0		& 0.9		& 0.9	 	& 0.9		& 1.0		& 0.9	& 0.9 	& 0.9		\\
\hline
Total	Events (NaNu)	& 10		& 10		& 30	& 10		& 7		& 7	 	& 20	& 5		\\
\hline
Total	Events (Super-NaNu)	& 60		& 60		& 180	& 45		& 40		& 40	 	& 115	& 30		\\
\hline
\end{tabular}
\caption{Overview of various tau decay channels including their branching ratio (BR) together with the efficiencies of various selection and identification criteria. The last two rows indicated the expected number of reconstructed $\nu _\tau$ events within NaNu and Super-NaNu.}
\label{tab:neutrinoreco}
\end{table*}

\section{Expected Physics Reach}

\subsection{Muon Neutrino Physics: Cross Sections, Deep Inelastic Scattering and Charm Production}

The expected statistics of the Mu-NaNu detector, operated in 2025 and 2026 together with the NA62 experiment yields, allows for an inclusive as well as differential cross-section measurement vs. the neutrino-energy between 10 and 50 GeV. Taking into account acceptance and reconstruction effects, we expect to reconstruct about 500 muon-neutrino and 100 anti-muon neutrino interactions.  The expected statistical uncertainty on the inclusive cross-section is therefore between 5\% and 10\%. 

The expected number of reconstructed $\nu_\mu$ and $\bar \nu_\mu$ interactions at the baseline NaNu detector is 25000 and 5000, respectively, requiring a minimal muon momentum of 5 GeV. Assuming additionally a minimal hadronic energy of the recoil system of 10 GeV, to allow for a sufficiently precise reconstruction, the numbers reduce by another 40\%. The inclusive cross section measurements are therefore non statistically limited, and differential cross-sections vs. Bjorken $x$ in range of 0.01-0.8 and the momentum transfer $Q^2$ in a range of 10 to 50 GeV can be measured. Assuming a two-dimensional binning of 5 bins in $x$ and $Q^2$ with similar statistics in each bin, the expected statistical uncertainties range between 5\% and 10\% for $\nu_\mu$ and $\bar \nu_\mu$ interactions, respectively. Those measurements would be an important consistency test of existing neutrino data in the context of global fits of parton density functions \cite{Candido:2023utz}. In particular, the energy of the hadronic recoil system $W^2$ can be measured and hence a separation between DIS, resonance- and scattering events with low momentum transfer could be distinguished. The study of those different processes at low energies would provide important information for upcoming neutrino experiments such as DUNE.

Also interestingly is the possibility to search for charm production in neutrino scattering events. This process is extremely sensitive to the s-quark content of nucleons, which plays a crucial role for several high precision measurements at the LHC such as the W boson mass \cite{ATLAS:2017rzl}. The identification of charm mesons can either be performed in the emulsion target or via their muonic decay channels. In the latter case, the full reconstruction can be performed using the active detector components and no reconstruction within the emulsion is required. The \textsc{Genie} program predicts about 4-5\% of charm meson production in muon-neutrino DIS scattering events. Taking acceptance and reconstruction efficiencies as well as minimal momentum requirements into account, we expect about 150 identified charm meson candidates which are identified in a two muon final state. This number can be increased by a factor of three to six relying on a charm-meson identification in the passive emulsion target, similar to the CHORUS experiment \cite{CHORUS:2005nog}, for the baseline NaNu and the Super-NaNu detectors respectively. The expected event yields of neutrino-induced charm production within NaNu is therefore at a similar level as previously achieved within CHORUS. It should be noted that several precision measurements which are used as input for global fits of parton distribution functions are in significant tension with each other. This is typically an indication that the stated systematic uncertainties have been largely underestimated. In this context, the charm measurements at NaNu would provide an important and independent test of the previously collected datasets on neutrino-induced charm production. 

\subsection{Tau Neutrino Physics: Discovery of Anti-Tau Neutrinos, Cross Sections, Structure Functions, Anomalous Magnetic Moment}

The number of identified  $\nu_\tau$ and $\bar \nu_\tau$ interactions exceeds the previous statistics by a factor 10 and 60 for NaNu and Super Nanu respectively, allowing in principle for a $\bar \nu_\tau$ observation during the first year of data taking within the baseline Nanu as well as the Super-Nanu setup. Beyond the first experimental observation of anti-tau neutrinos, we expect to measure the inclusive cross section of  $\nu_\tau$ and $\bar \nu_\tau$ interactions with a statistical precision of 10\% at the baseline NaNu detector and 5\% at Super-NaNu. The latter would also allow for differential cross-section measurement for both neutrino flavors in two energy regimes. It should be noted, that the expected systematic uncertainties on the measurement are also on the 5\% level. Hence an increase of statistics would not necessarily leads to an improved precision. 

The cross-section measurement of $\nu_\tau$ and $\bar \nu_\tau$ interactions can also be used to constrain the $F_4$ and $F_5$ structure functions \cite{Albright:1974ts} for the first time. The contribution to the neutrino cross-sections of $F_4$ and $F_5$ are negligible in electron and muon neutrino interactions due to light charged lepton mass, however, become sizable for tau and anti-tau neutrino interactions. The effect of $F_4$ and $F_5$ on the $\nu_\tau$ cross-section is about 30\% at $E_\nu$=20 GeV and decreases for higher energies. The effect for $\bar \nu_\tau$ interactions is even larger. Given that the expected $\nu_\tau$ energies are in the same range where the maximal effect is expected, a significant first constraint on $F_4$ and $F_5$ will be possible within NaNu. Super-NaNu will even be able to test the energy dependence as well as study the difference between $\nu_\tau$ and $\bar \nu_\tau$ interactions.

Similarly to the determination of the upper limit for the magnetic moment of the tau neutrino by the DONUT Collaboration~\cite{DONUT:2000fbd,DONUT:2001zvi}, a study on the $\nu_\tau$ magnetic moment can be performed at NaNu. It is reasonable to assume similar systematic uncertainties but a significantly improved statistics. A detailed study is currently ingoing, however it seems promising to lower the upper limit on the magnetic moment of the tau neutrino with NaNu and even more with Super-NaNu.

\subsection{Tests of Lepton Universality and Searches for new physics}

The NaNu datasets provides a unique opportunity to test lepton universality. The neutrino interaction cross-sections are expected to be systematically limited for electron and muon neutrinos. While the tau neutrino interaction cross sections are statistically limited for the baseline NaNu detector, they will be also systematically limited for Super-NaNu. By looking at cross section ratios, several systematic uncertainties will cancel, e.g. partly the uncertainties on the initial neutrino flux. 

The integration of the NaNu detector system, in particular its active components, in the Shadows experiment would allow to extend the search for long-lived particles. Maybe more interestingly, the emulsion detector can be used for the direct search of signatures of light bosons. Detailed studies are ongoing.

\section{Estimated Costs}

The existing dipole magnet (MNP 22/B) is the baseline choice for the magnet-system of the NaNu experiment. While it should be fully functional, it was not in active use in the past years and certain repair and refurbishing costs might arise; additionally a suitable power converter should be available on site to operate the magnet. The weight of the emulsion detector system requires a dedicated support frame, which allows an easy access for replacing the emulsion layers. Based on the support frame costs of the FASER experiment, 30~k\euro{} should be sufficient to build the support frame. 

The costs of the first phase of NaNu, i.e. the Mu-NaNu detector are summarized in Table \ref{tab:Mu-NaNucosts}. The largest costs are expected for the micromegas-based tracking chambers, which can be produced by the CERN workshops with a preliminary cost estimate of 90~k\euro{} in total. The readout system will be based on the VMM-2 chipset~\cite{Iakovidis:2018nwp}, yielding overall costs of 120~k\euro{}. The purchase of 4.8t tungsten as the passive target material has expected costs of 150 \euro{}, which is significantly cheaper than the thin tungsten plates required for the emulsion detector. The scintillation plates in the target as well as those used for the muon veto have estimated costs of 60 \euro{}, including a SiPM readout system. Additional costs of 100 \euro{} for further infrastructure investments (gas, HV, computing) are foreseen. In total, Mu-NaNu could be built for 650 k\euro{}. Given the simplicity of the design, the experience with all sub-detector components as well as the good availibility of the required components, a construction time in less than one year seems not unrealisitic.  

\begin{table}[ht]
\centering
\begin{tabular}{l |r } 
\hline  
Magnet Refurbishment						& 100 k\euro{}	\\
\hline  
Support Structure							& 30 k\euro{}	\\
\hline
Tungsten Bricks (4800 kg) 					& 150 k\euro{}	\\
\hline
Micromegas Tracking Stations (9x)				& 90 k\euro{}	\\
Micromegas Readout System (VMM chip) 		& 120 k\euro{}	\\
\hline
Plastic Scintillators					 		& 50 k\euro{}	\\
SiPM Scintillator Readout				 		& 10 k\euro{}	\\
\hline
Gas-System								& 30	k\euro{}	\\
High-Voltage Supply							& 40 k\euro{}	\\
Computing								& 30 k\euro{}	\\
\hline
{\bf Total }									&\textbf{650 k\euro{}}	\\
\hline
\end{tabular}
\caption{Overview of the expected costs for the Mu-NaNu Experiment. \label{tab:Mu-NaNucosts}} 
\end{table}

As already mentioned, the Mu-NaNu detector is a perfect starting point for the later baseline NaNu system, as only one half of the active detector has be removed and replaced by the emulsion detector. The expected costs of the emulsion detector, including gel and film as well as additional infrastructure such as chemicals and tooling are estimated based on the costs of the FASER$\nu$ experiment~\cite{FASER:2019dxq} and scaled to the larger emulsion detector volume. The costs of the tungsten plates, which will be used as absorber material, is conservatively estimated using current market prices for thin tungsten plates. The corresponding costs are summarized in \autoref{tab:costs}. Largest costs driver are the tungsten plates as well as the yearly costs of the emulsion gel and film, yielding total costs of 1,850 k\euro{}.

The cost of Super-NaNu scales nearly by a factor of two compared to the baseline NaNu detector system as the size of the emulsion detector doubles. In addition, it is likely to be necessary to install a dedicated magnetized muon shielding to reduce the expected muon background rate below $1\times10^6$/cm$^2$. Preliminary simulation studies suggest that this is possible, however, a first reference measurement of the actual muon background flux is still to be performed. The costs of a dedicate shielding would amount to 100 to 200 k\euro{}.

\begin{table}[ht]
\centering
\begin{tabular}{l |r } 
\hline
\textit{Investments of for Emulsion Detector}		&			\\
Infrastructure (chemicals, tools, racks)		 	& 50 k\euro{}	\\
Tungsten Plates (2400 kg) 					& 600 k\euro{}	\\
\hline
\textit{Yearly Costs of the Emulsion Detector}		&			\\
Emulsion Detector (Gel + Film) 					& 1000 k\euro{}	\\
Emulsion Film Production Costs 				& 200 k\euro{}	\\
\hline
{\bf Total }									&\textbf{1,850 k\euro{}}	\\
\hline
\end{tabular}
\caption{Overview of the expected costs for the NaNu Experiment. The costs for the emulsion detector as well as the production costs cover one year of running, allowing for a first observation of anti-tau neutrino events. This table assumes that the Mu-NaNu Experiment has been already established in front of NA62. \label{tab:costs}} 
\end{table}

\section{Summary}

We proposed a cost effective experiment to study neutrino interactions at the SPS collider in an energy range of 10 to 60~GeV. While first version of NaNu, consisting only of active detector components with a continuous readout could be already installed in near future in front of the NA62 experiment, the baseline version of NaNu is foreseen to be realized in the North Area of CERN between the future SHADOWS and HIKE Experiments. The first experimental distinction of tau and anti-tau neutrinos can be already achieved with the first year of data taking. Furthermore, new upper bounds on the anomalous magnetic moment of tau neutrino are expected as well as precise neutrino cross-section measurements of all flavors. The study of muon neutrino interactions at NaNu will improve our understanding of deep inelastic scattering in the neutrino sector, in particular to charm-meson final states. 

This paper should mainly serve as a proof of concept and more detailed studies on the physics reach as well as the final detector concept are necessary. Most performance estimates have the be treated as preliminary and the reported numbers have significant modeling uncertainties associated. However, it becomes evident, that tau neutrino physics can be successfully conducted also off-beam axis in parallel to the operation of the SHADOWS experiment. 

More investments on the additional (and dedicated) muon shielding and a larger emulsion detector would allow to realize an advanced NaNu detector option that would increase the expected number of reconstructed tau neutrino interactions by a factor of five, thus allowing for a significant improvement on several measurements. It should be noted that the envisioned energy range is complementary to the neutrino program of FASER-2 and even a common development of baseline technology could be envisioned. 

\section*{Acknowledgement}
We thank Alfons Weber for the fruitful discussions during the preparation of this document, Sam Zeller for his help on the neutrino-interaction cross-sections, Akitaka Ariga for his input on the emulsion detector system as well as to Juan Rojo for his input on the study of deep inelastic scattering processes. Special thanks also go to Florian Stummer for the muon background studies as well as to Giovanni de Lellis, who pointed us to an important mistake in our first estimate of the expected neutrino fluxes. This proposal would not have been possible without the support of the PRISMA$^+$ Cluster of Excellence as well as the ERC Grant Light@LHC. 

\bibliographystyle{elsarticle-num}
\bibliography{./Bibliography}

\section*{Appendix}

An overview of the expected reduction factors on the number of expected neutrino interactions in different setups of the NaNu Experiment compared to the SND@Ship experiment is given in Table \ref{tab:reductionfactors}. The overview on physics reach of the baseline NaNu as well as the Super-NaNu setup is shown in Table \ref{tab:physicsreach}.

\begin{table}[htb]
\scriptsize
\centering
\begin{tabular}{l | c c c | c}
\hline
Setup			&Lumi./		& Target-	& Position & Total \\
				&POT		& Mass	& 		 & 		 \\
\hline
Mu-NaNu ($\nu_\mu$) 			& 0.005 & 0.6& $\approx$0.1	& 0.0003 \\
\hline
Baseline NaNu ($\nu_\mu$) 		& 0.25 & 0.6	& $\approx$0.1	& 0.015 \\
Baseline NaNu ($\nu_e, \nu_\tau$) 	& 0.25 & 0.3	& $\approx$0.05	& 0.004 \\
\hline
Super-NaNu  ($\nu_\mu$)			& 0.25 & 0.6	& $\approx$0.1	& 0.015 \\
Super-NaNu ($\nu_e, \nu_\tau$) 	& 0.25 & 0.3	& 0.1-0.15& 0.01 \\
\hline
\end{tabular}
\caption{Overview of reduction factors on the number of expected neutrino interactions in different setups of the NaNu Experiment compared to the SND@Ship experiment.}
\label{tab:reductionfactors}
\end{table}

\begin{table}[htb]
\scriptsize
\centering
\begin{tabular}{l | c c }
\hline
Physics Reach				& NaNu	& Super-NaNu \\
\hline
Expected $\nu_\tau / \bar \nu_\tau$	& 60/40	& 340/220	\\
$\bar \nu_\tau$ observation		& yes		& yes		\\
$\nu_\tau$ magnetic moment		& yes		& yes		\\
\hline
Lepton-flavor universality			& yes		& yes		\\
DIS structure function $F_4, F_5$	& yes		& yes		\\
Sea quark PDFs, charm production	& yes		& yes		\\
Input for PDF Fits				& yes		& yes		\\
Cross sections for Future Experiments	& yes		& yes		\\
\hline

\end{tabular}
\caption{Summary of physics reach of the baseline NaNu and Super NaNu Detectors.}
\label{tab:physicsreach}
\end{table}

\end{document}

%% file: Tools/command_list.tex
\newcommand{\bi}{\begin{itemize}}
\newcommand{\ei}{\end{itemize}}

\newcommand{\ben}{\begin{enumerate}}
\newcommand{\een}{\end{enumerate}} 

\newcommand{\bt}[1]{\begin{table}[tb]\begin{tabular}{#1} \hline\hline  \\[-1.0em]}
\newcommand{\et}[2]{\hline\hline \end{tabular} \caption{#1} \label{#2} \end{table}}

\newcommand{\be}{\begin{equation}}
\newcommand{\ee}{\end{equation}}

\newcommand{\bea}{\begin{eqnarray}}
\newcommand{\eea}{\end{eqnarray}}

\newcommand{\bc}{}

\newcommand{\mev}{\ensuremath{\mathrm{\,Me\kern -0.1em V}}\xspace}
\newcommand{\gev}{\ensuremath{\mathrm{\,Ge\kern -0.1em V}}\xspace}



%% file: author_list.tex

\author{Friedemann Neuhaus \and Matthias Schott$^*$ \and Chen Wang \and Rainer Wanke}
\institute{Institute of Physics and PRISMA$^+$ Cluster of Excellence, Johannes Gutenberg University, Mainz, Germany}
\institute{{\tiny$^*$ corresponding author}}

%% file: main.bbl
\begin{thebibliography}{10}
\expandafter\ifx\csname url\endcsname\relax
  \def\url#1{\texttt{#1}}\fi
\expandafter\ifx\csname urlprefix\endcsname\relax\def\urlprefix{URL }\fi
\expandafter\ifx\csname href\endcsname\relax
  \def\href#1#2{#2} \def\path#1{#1}\fi

\bibitem{FASER:2018eoc}
A.~Ariga, et~al., {FASER\textquoteright{}s physics reach for long-lived
  particles}, Phys. Rev. D 99~(9) (2019) 095011.
\newblock \href {http://arxiv.org/abs/1811.12522} {\path{arXiv:1811.12522}},
  \href {http://dx.doi.org/10.1103/PhysRevD.99.095011}
  {\path{doi:10.1103/PhysRevD.99.095011}}.

\bibitem{Ahdida:2750060}
C.~Ahdida, et~al., {SND@LHC - Scattering and Neutrino Detector at the LHC},
  Tech. Rep. \href{https://cds.cern.ch/record/2750060}{CERN-LHCC-2021-003,
  LHCC-P-016}, CERN, Geneva (2021).

\bibitem{DONUT:2000fbd}
K.~Kodama, et~al., {Observation of tau neutrino interactions}, Phys. Lett. B
  504 (2001) 218--224.
\newblock \href {http://arxiv.org/abs/hep-ex/0012035}
  {\path{arXiv:hep-ex/0012035}}, \href
  {http://dx.doi.org/10.1016/S0370-2693(01)00307-0}
  {\path{doi:10.1016/S0370-2693(01)00307-0}}.

\bibitem{OPERA:2010pne}
N.~Agafonova, et~al., {Observation of a first $\nu_\tau$ candidate in the OPERA
  experiment in the CNGS beam}, Phys. Lett. B 691 (2010) 138--145.
\newblock \href {http://arxiv.org/abs/1006.1623} {\path{arXiv:1006.1623}},
  \href {http://dx.doi.org/10.1016/j.physletb.2010.06.022}
  {\path{doi:10.1016/j.physletb.2010.06.022}}.

\bibitem{Workman:2022ynf}
R.~L. Workman, Others, {Review of Particle Physics}, PTEP 2022 (2022) 083C01.
\newblock \href {http://dx.doi.org/10.1093/ptep/ptac097}
  {\path{doi:10.1093/ptep/ptac097}}.

\bibitem{SHiP:2015vad}
M.~Anelli, et~al., {A facility to Search for Hidden Particles (SHiP) at the
  CERN SPS}~(\href{https://cds.cern.ch/record/2007512}{CERN-SPSC-2015-016,
  SPSC-P-350}).
\newblock \href {http://arxiv.org/abs/1504.04956} {\path{arXiv:1504.04956}}.

\bibitem{Ahdida:2654870}
C.~Ahdida, et~al., {SHiP Experiment - Progress Report}, Tech. Rep.
  \href{https://cds.cern.ch/record/2654870}{CERN-SPSC-2019-010, SPSC-SR-248},
  CERN, Geneva (2019).

\bibitem{Bai:2018xum}
W.~Bai, M.~H. Reno, {Prompt neutrinos and intrinsic charm at SHiP}, JHEP 02
  (2019) 077.
\newblock \href {http://arxiv.org/abs/1807.02746} {\path{arXiv:1807.02746}},
  \href {http://dx.doi.org/10.1007/JHEP02(2019)077}
  {\path{doi:10.1007/JHEP02(2019)077}}.

\bibitem{Baldini:2799412}
W.~Baldini, et~al., {SHADOWS (Search for Hidden And Dark Objects With the SPS):
  Expression of Interest}, Tech. Rep.
  \href{https://cds.cern.ch/record/2799412}{CERN-SPSC-2022-006, SPSC-EOI-022},
  CERN, Geneva (2022).

\bibitem{HIKELoI}
E.~Cortina~Gil, et~al., {HIKE, the High Intensity Kaon Experiments at the CERN
  SPS: Letter of Intent}, Tech. Rep. to be published.

\bibitem{FASER:2019dxq}
H.~Abreu, et~al., {Detecting and Studying High-Energy Collider Neutrinos with
  FASER at the LHC}, Eur. Phys. J. C 80~(1) (2020) 61.
\newblock \href {http://arxiv.org/abs/1908.02310} {\path{arXiv:1908.02310}},
  \href {http://dx.doi.org/10.1140/epjc/s10052-020-7631-5}
  {\path{doi:10.1140/epjc/s10052-020-7631-5}}.

\bibitem{Brickwedde:2016lhu}
B.~Brickwedde, A.~D\"udder, M.~Schott, E.~Yildirim, {Design, Construction and
  Performance Tests of a Prototype MicroMegas Chamber with Two Readout Planes
  in a Common Gas Volume}, Nucl. Instrum. Meth. A 864 (2017) 1--6.
\newblock \href {http://arxiv.org/abs/1610.09539} {\path{arXiv:1610.09539}},
  \href {http://dx.doi.org/10.1016/j.nima.2017.04.010}
  {\path{doi:10.1016/j.nima.2017.04.010}}.

\bibitem{Bierlich:2022pfr}
C.~Bierlich, et~al., {A comprehensive guide to the physics and usage of PYTHIA
  8.3}\href {http://arxiv.org/abs/2203.11601} {\path{arXiv:2203.11601}}.

\bibitem{Andreopoulos:2009rq}
C.~Andreopoulos, et~al., {The GENIE Neutrino Monte Carlo Generator}, Nucl.
  Instrum. Meth. A 614 (2010) 87--104.
\newblock \href {http://arxiv.org/abs/0905.2517} {\path{arXiv:0905.2517}},
  \href {http://dx.doi.org/10.1016/j.nima.2009.12.009}
  {\path{doi:10.1016/j.nima.2009.12.009}}.

\bibitem{Alekhin:2015byh}
S.~Alekhin, et~al., {A facility to Search for Hidden Particles at the CERN SPS:
  the SHiP physics case}, Rept. Prog. Phys. 79~(12) (2016) 124201.
\newblock \href {http://arxiv.org/abs/1504.04855} {\path{arXiv:1504.04855}},
  \href {http://dx.doi.org/10.1088/0034-4885/79/12/124201}
  {\path{doi:10.1088/0034-4885/79/12/124201}}.

\bibitem{Agostinelli:2002hh}
S.~Agostinelli, et~al., {GEANT4--a simulation toolkit}, Nucl. Instrum. Meth. A
  506 (2003) 250--303.
\newblock \href {http://dx.doi.org/10.1016/S0168-9002(03)01368-8}
  {\path{doi:10.1016/S0168-9002(03)01368-8}}.

\bibitem{Candido:2023utz}
A.~Candido, A.~Garcia, G.~Magni, T.~Rabemananjara, J.~Rojo, R.~Stegeman,
  {Neutrino Structure Functions from GeV to EeV Energies}\href
  {http://arxiv.org/abs/2302.08527} {\path{arXiv:2302.08527}}.

\bibitem{ATLAS:2017rzl}
M.~Aaboud, et~al., {Measurement of the $W$-boson mass in pp collisions at
  $\sqrt{s}=7$ TeV with the ATLAS detector}, Eur. Phys. J. C 78~(2) (2018) 110,
  [Erratum: Eur.Phys.J.C 78, 898 (2018)].
\newblock \href {http://arxiv.org/abs/1701.07240} {\path{arXiv:1701.07240}},
  \href {http://dx.doi.org/10.1140/epjc/s10052-017-5475-4}
  {\path{doi:10.1140/epjc/s10052-017-5475-4}}.

\bibitem{CHORUS:2005nog}
A.~Kayis-Topaksu, et~al., {Measurement of topological muonic branching ratios
  of charmed hadrons produced in neutrino-induced charged-current
  interactions}, Phys. Lett. B 626 (2005) 24--34.
\newblock \href {http://dx.doi.org/10.1016/j.physletb.2005.08.082}
  {\path{doi:10.1016/j.physletb.2005.08.082}}.

\bibitem{Albright:1974ts}
C.~H. Albright, C.~Jarlskog, {Neutrino Production of m+ and e+ Heavy Leptons.
  1.}, Nucl. Phys. B 84 (1975) 467--492.
\newblock \href {http://dx.doi.org/10.1016/0550-3213(75)90318-1}
  {\path{doi:10.1016/0550-3213(75)90318-1}}.

\bibitem{DONUT:2001zvi}
R.~Schwienhorst, et~al., {A New upper limit for the tau - neutrino magnetic
  moment}, Phys. Lett. B 513 (2001) 23--29.
\newblock \href {http://arxiv.org/abs/hep-ex/0102026}
  {\path{arXiv:hep-ex/0102026}}, \href
  {http://dx.doi.org/10.1016/S0370-2693(01)00746-8}
  {\path{doi:10.1016/S0370-2693(01)00746-8}}.

\bibitem{Iakovidis:2018nwp}
G.~Iakovidis, {VMM - An ASIC for micropattern detectors}, EPJ Web Conf. 174
  (2018) 07001.
\newblock \href {http://dx.doi.org/10.1051/epjconf/201817407001}
  {\path{doi:10.1051/epjconf/201817407001}}.

\end{thebibliography}
